\documentclass[%
 reprint,
 amsmath,amssymb,
 aps,
pra,
]{revtex4-2}

\usepackage{graphicx}
\usepackage{dcolumn}
\usepackage{bm}
\usepackage{braket}
\usepackage{url}

\newcommand{\e}{\text{e}}

\expandafter\let\csname equation*\endcsname=\relax 
\expandafter\let\csname endequation*\endcsname=\relax 

\begin{document}

\preprint{APS/123-QED}

\title{Transit effects for non-linear index measurement in hot atomic vapors.}

\author{Tangui Aladjidi$^*$}
 \author{Murad Abuzarli}%
\homepage{These authors contributed equally to this work.}
 \author{Guillaume Brochier}
  \author{Tom Bienaimé}
  \altaffiliation[Now at ]{ISIS (UMR 7006), University of Strasbourg and CNRS, 67000 Strasbourg, France}
   \author{Thomas Picot}
    \author{Alberto Bramati}
     \author{Quentin Glorieux}
 \email{quentin.glorieux@lkb.upmc.fr}
\affiliation{Laboratoire Kastler Brossel, Sorbonne Universit\'{e}, ENS-Universit\'{e} PSL, Coll\`{e}ge de France, CNRS, 4 place Jussieu, 75252 Paris Cedex 05, France}



\date{\today}

\begin{abstract}
Hot atomic vapors are widely used in non-linear and quantum optics due to their large Kerr non-linearity.
While the linear refractive index and the transmission are precisely measured and well modeled theoretically, similar characterization remains partial for the $\chi^{(3)}$ non-linear part of the susceptibility. 
In this work, we present a set of tools to measure and estimate numerically the non-linear index of hot atomic vapors both in the steady state and during the transient response of the medium.
We apply these techniques for the characterization of a hot vapor of rubidium and we evidence the critical role played by transit effects, due to finite beam sizes, in the measurement of the non-linear index.
\end{abstract}

\maketitle


\section{Introduction}
When light propagates in a medium, it acquires a phase which depends on the refractive index of the medium $n$.
At low intensity, the response of the medium is usually linear and the accumulated phase does not depend on the intensity. 
On the contrary, at large intensity, the medium starts to respond non-linearly and the total refractive index depends on the light intensity.
Interestingly, for alkali vapors near resonance, this effect becomes significant for intensities as low as mW/cm$^2$. 
More intriguing still, in atomic vapors, the effect depends not only on the intensity of the beam, but also on its size.
This non-linear behavior is at the origin of a broad range of experiments in non-linear and quantum optics
\cite{RevModPhys.82.1041,julsgaard2004experimental,katz2018light,glorieux2011quantum,glorieux2012temporally,glorieux2012generation}.
Moreover, hot atomic vapors have emerged, recently, as a medium of choice for paraxial fluids of light \cite{larre2015propagation, fontaine2018observation, vsantic2018nonequilibrium, abuzarli2021blast, bienaime2021quantitative, steinhauer2021analogue}. 
In this approach, the self-defocusing due to the non-linear refractive index is interpreted as a repulsive two-body interaction between the photons, and therefore an exhaustive characterization of the non-linearity is required for quantitative analyses.
While the linear refractive index (both the real and imaginary part of the susceptibility) has been studied in detail for hot atomic vapors \cite{siddons2008absolute,siddons2009off,weller2011absolute}, a comprehensive study of the non-linear index is still missing.\\

Various techniques to measure the non-linear index of a medium have been proposed including methods based on beam-deflection  \cite{mcconville_measurement_2005,purves_refractive_2004,rasouli_nonlinear_2012}, z-scan transmission  \cite{sheik-bahae_high-sensitivity_1989,santos_measurement_2019,araujo_measurement_2013} and ring patterns in far-field imaging \cite{boughdad_anisotropic_2019,vsantic2018nonequilibrium}.
However, all these techniques suffer from limitations either in the applicability, the simplicity of implementation or the precision.
For instance, the most commonly used method is the so called "z-scan" technique \cite{sheik-bahae_high-sensitivity_1989, McCormick:03, mccormick_saturable_2004,wang2020measurement,dos2021theoretical}.
This technique consists in the measurement of the normalized transmission through a pinhole aperture placed in the far-field of the sample as a function of its position with respect to the waist of a focused Gaussian beam.
From a practical perspective, this approach requires multiple transmission measurements to access the index variation at a given laser frequency and cannot provide information about the role of the beam size on the non-linear index since the sample needs to move through the focus of the beam.
Moreover, this technique requires to work with a thin sample, in order to minimize beam extinction, which drastically reduces its applicability for atomic vapor cells.\\

Another approach, mainly used for paraxial fluids of light characterization \cite{fontaine2018observation,vsantic2018nonequilibrium,bienaime2021quantitative,boughdad_anisotropic_2019}, relies on imaging the intensity dependent annular pattern in the far-field of the sample \cite{durbin_laser-induced_1981}.
The origin of the pattern resides in multi-wave interference of distinct points within the beam having the same deflection induced by non-linearity.
The number of rings increases linearly with the accumulated non-linear phase shift and allows to retrieve directly the non-linear index.
However, at large intensities (corresponding to typically 10 rings or more), the interference visibility lowers and the self defocusing also affects the non-linear phase reconstruction, limiting this method to low intensity and wide beams.\\

A different class of measurements, based on wavefront measurement using interferometric techniques, lifts most of these limitations. 
The general idea is to place the non-linear medium in one arm of an interferometer (typically a Mach–Zehnder interferometer) and use phase retrieval algorithm to reconstruct the accumulated phase from the interferogram \cite{boudebs_third-order_2001, RODRIGUEZ2005453,dancus_single_2013}. 
This method is more precise than the z-scan or ring patterns approaches, easy to implement thanks to a single shot approach, and much more flexible since it allows for thick samples, time resolved measurements and spatial resolution.
While similar methods have been used for solid state systems \cite{olbright1986interferometric}, this technique has not yet been applied for hot atomic vapor non-linear index characterization.\\
\noindent In this work, our goal is three-fold: 
\begin{itemize}
    \item We present a detailed procedure to use interferometric phase retrieval for non-linear index measurement in thick samples, with two complementary analysis methods and a complete process automation, which are applicable in all non-linear media.
    \item We apply this technique to a hot atomic vapor of rubidium and evidence the crucial impact of the transverse beam size on the non-linear index.
    \item We propose an interpretation of this effect as a direct consequence of the transit time of warm atoms inside the laser beam and we provide Monte-Carlo simulations of optical Bloch equations to validate this hypothesis.
\end{itemize}

\section{Non-linear phase measurements}
The Taylor expansion of the electric susceptibility in a centro-symmetric medium only includes odd terms ($\chi^{(1)}, \chi^{(3)}, \chi^{(5)}...$) due to symmetry.
At the lowest order of approximation taking into account the non-linear response of the medium, this expansion can be truncated after the $\chi^{(3)}$ term and the refractive index $n$ takes the form $n=n_0+n_2 I$, where $n_0$ is the linear index and $\Delta n= n_2 I$ is the non-linear index i.e. the product of the non-linear coefficient $n_2 $ by the intensity $I$.
This approximation is widely used, for example to describe fluids of light with contact interactions \cite{fontaine2018observation}.
However, at larger intensity, higher order terms should not be neglected and the full expansion can be rewritten in terms of refractive index as:
\begin{equation}\label{eq:Dn}
    n= n_0 + \Delta n = n_0+ n_{2}\frac{\tilde{I}}{1+\frac{\tilde{I}}{I_\text{S}}}.
\end{equation}
The non-linear index $\Delta n$ is modified to take into account the saturation of the medium non-linearity $I_\text{S}$ \cite{mccormick_saturable_2004}, and the absorption through the cell $\tilde{I}=I\frac{1-e^{-\alpha L}}{\alpha L}$ with $\alpha$ is the linear absorption coefficient. Typical values for $I_s$ are in the range of 10 to 500 W/cm$^2$ depending on waist size (see section 3.).
In this section, we present an interferometric technique and two complementary analysis methods to measure $\Delta n$ and extract the non-linear coefficient $n_2$ and and the saturation $I_\text{S}$.
While these methods are general and can be applied for a wide variety of non-linear media, we focus, here, on the non-linear index of a hot atomic vapor of rubidium 87 near atomic resonance on the D2 line. The detuning to resonance is denoted $\Delta=\nu-\nu_0$, where $\nu_0$=380.284 THz \cite{Steck_numbers} is the $^{87}$Rb D2 line frequency and $\nu$ is the laser frequency.

\subsection{Non-linear Mach–Zehnder interferometer}
\begin{figure}[h]
    \centering
    \includegraphics[width=\linewidth]{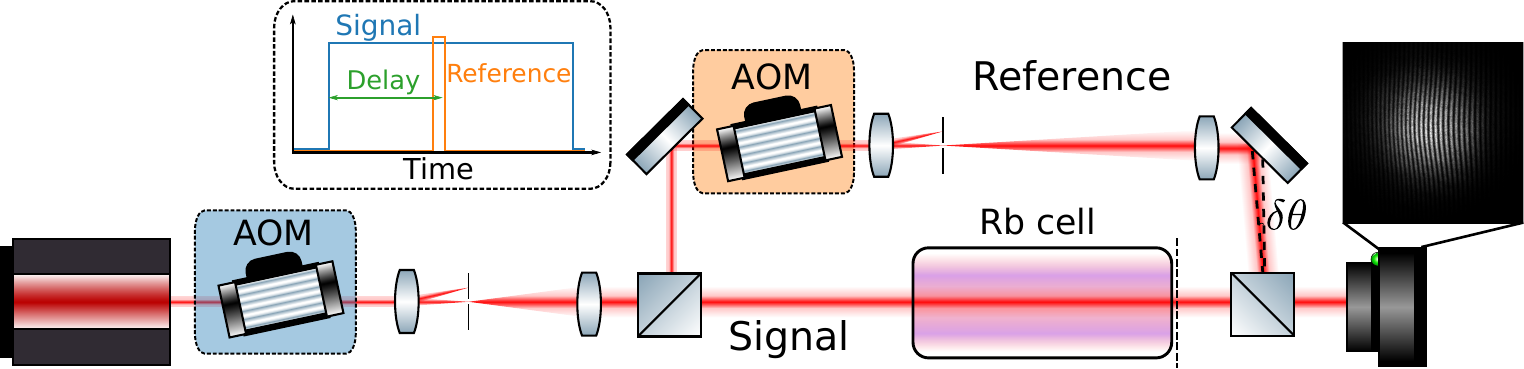}
    \caption{Mach-Zehnder interferometer used for the non-linear phase measurement. 
    A laser source (a fiber-amplified frequency-doubled diode laser) is split in two arms (signal and reference) and recombined with angle $\delta\theta$ in order to modulate the signal through interferences.
    Image of a typical interference pattern at the output of the cell is shown above the camera. The image plane of the camera is shown by the dashed line after the cell.
    Inside dashed boxes, we present the modification required for time-resolved operation. AOM are acousto-optic modulators used for pulsing the beams.
    The atomic response is sampled temporally by timing the delay between the signal and reference pulses as shown in the dashed box. }
    \label{fig:setup}
\end{figure}

Interferometric techniques rely on measuring the local phase difference between a reference beam and a beam that has interacted with a non-linear medium.
In our experiment, this is done by inserting the non-linear medium in one arm of a Mach–Zehnder interferometer (see Fig. \ref{fig:setup}). 
In order to achieve a high spatial resolution, we add a tilt ($\delta \theta$) to the reference beam that leads to the fringe pattern shown in the inset of Fig.~\ref{fig:setup}.
Note that, on the same figure, we also present a modification of the setup (within dashed boxes) required for time-resolved operation described in section~\ref{section:time-resolved}.
The output of the interferometer is  recorded either on a camera or on a photodiode, depending on the analysis method.
In the following, we give a detailed description of two complementary analysis procedures for extracting the non-linear index.

\subsection{Fourier filtering phase retrieval method}
\paragraph{Fourier analysis}
\begin{figure}[h]
    \centering
    \includegraphics[width=0.9\textwidth]{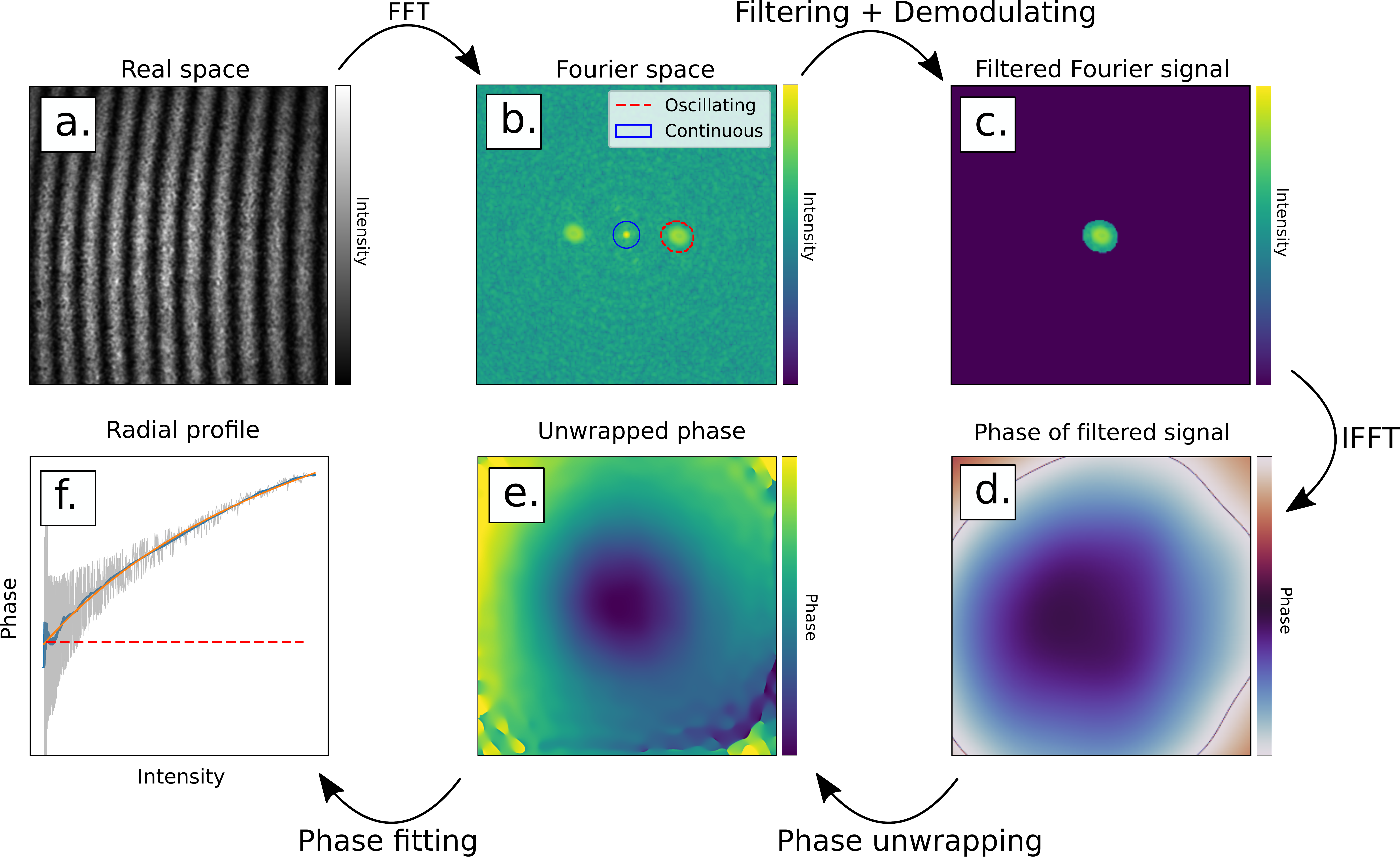}
  \caption{The interferogram (a) is first Fourier transformed (b), then the satellite peak is automatically detected using an image recognition routine, then the filtered peak is centered in the Fourier plane for demodulation (c). 
    Finally, we inverse transform the signal (d) and unwrap the resulting phase (e).
    In panel (f), the geometrical phase offset is obtained from the fit indicated by orange line. The red dashed line is the phase offset and the grey line represents the phase values of the unwrapped phase.}
    \label{fig:intro}
\end{figure}


When interfering with the signal beam, the reference beam slices a cut of the signal wavefront at a fixed angle. 
The intensity detected on the camera as function of $\mathbf{r}=(x,y)$ is given by:

\begin{widetext}
\begin{equation}
\begin{split}
    I_\text{camera} (\mathbf{r}) & \propto | \mathcal{E}_\text{s}(\mathbf{r}) \, \e^{i(\mathbf{k}_\text{s} \mathbf{r} + \varphi(\mathbf{r}))} + \mathcal{E}_\text{r}(\mathbf{r}) \, \e^{i\mathbf{k}_\text{r} \mathbf{r}}| ^2 = | \mathcal{E}_\text{s}(\mathbf{r}) \, \e^{i \varphi(\mathbf{r})} + \mathcal{E}_\text{r}(\mathbf{r}) \, \e^{i \textbf{k}_\perp \textbf{r}_\perp}|^2 \\
    & = \underbrace{\vphantom{\left( \e^i \right)} I_\text{s} (\mathbf{r}) + I_\text{r} (\mathbf{r})}_{\text{continuous part}} 
    + \underbrace{\epsilon_0 c \, \Re \left( \mathcal{E}_\text{s}(\mathbf{r})  \mathcal{E}^{*}_\text{r}(\mathbf{r})  \e^{i(\varphi (\mathbf{r}) + \textbf{k}_\perp \textbf{r}_\perp)} \right)}_{\mathcal{V}(\mathbf{r})} ~,
\end{split}
\label{eq:intensity_mz_2}
\end{equation}
\end{widetext}

where we have used the fact that the angle between the two beams $\delta \theta$ is small and $\mathbf{k}_\perp = (k_x, k_y)$ is the projection on the image plane of the reference beam wavevector $\mathbf{k_r}$. 
$\varphi(\mathbf{r})$ is the phase accumulated by the signal beam, which corresponds to the nonlinear phase plus an offset. 
The two electric fields ($\mathcal{E}_\text{s}$ and $\mathcal{E}_\text{r}$) are linearly polarized in the same direction.

The off-axis contribution performs a shift in the Fourier space for the last term of equation \eqref{eq:intensity_mz_2}.
This can be interpreted as a spatial heterodyne detection, with the reference beam shifting the frequencies of the signal, and the demodulation being done numerically.
Phase information can therefore be obtained by filtering $\mathcal{V(\mathbf{r})}$ in the Fourier space. 
Taking the Fourier Transform of this expression yields:
\begin{equation}
\begin{split}
    \label{eq:tf_mz}
   & \tilde{I}_\text{camera} (\mathbf{k}) = \tilde{I}_\text{s} (\mathbf{k}) + \tilde{I}_\text{r} (\mathbf{k}) +\\
    &\epsilon_0 c \, \mathcal{F} \left[\mathcal{E}_\text{s}\e^{i\varphi (\mathbf{r})} \right] (\mathbf{k}) *( \mathcal{F} \left[\mathcal{E}_\text{r}\right] (\mathbf{k} - \mathbf{k}_\perp)  + \mathcal{F} \left[\mathcal{E}_\text{r}\right] (\mathbf{k} + \mathbf{k}_\perp) ) ~,
    \end{split}
\end{equation}

where $\tilde{\null}$ and $\mathcal{F}$ means the Fourier transform of a quantity and $*$ denotes a convolution product.
In practice, the reference beam has a gaussian profile much wider than the signal and the main effect of the convolution product is to shift the information on $\mathcal{E}_s e^{i\varphi(\textbf{k})}$ in two symmetrical satellite peaks on both sides of the continuous part.
After filtering out the continuous part, we compute the inverse Fourier Transform of one satellite peak that yields directly half of $\mathcal{V}(\mathbf{r})$. By taking the argument of this quantity, we recover the total phase $\Phi(\mathbf{r})$:
\begin{widetext}
\begin{equation}
    \label{eq:phi_tot}
    \Phi(\mathbf{r}) =\, \varphi (\mathbf{r}) + \mathbf{k}_\perp\cdot\mathbf{r}_\perp = \underbrace{\varphi_{\textsc{nl}}(\mathbf{r})}_{\text{nonlinear phase}} \,  + \, \overbrace{\varphi_0}^{\text{geometrical phase}}  + \, \underbrace{\vphantom{\varphi_{\textsc{nl}}} k_x x + k_y y}_{\text{~off-axis contribution}} \, .
\end{equation}
\end{widetext}
In order to retrieve the nonlinear phase from the total phase, we proceed in two steps:
\begin{itemize}
    \item the off-axis contribution is removed by measuring $\mathbf{k}_\perp$ through the position of a satellite peak. Demodulation is then readily done by shifting the peak back to the center of the Fourier plane;
    \item we get rid of the geometrical phase offset by fitting the phase as function of intensity by $\Delta n(I, I_S, b) = n_{2}\frac{I}{1+\frac{I}{I_s}} + b$. We can then extend the zero intensity behaviour using the offset b of the fit. This allows to eliminate any constant phase offset picked up during propagation.
\end{itemize}
We then retrieve the full 2D non-linear phase map $\varphi_{NL}(\textbf{r})$.
This process is summarized in Fig.~\ref{fig:intro}.

\paragraph{Automation of the satellite peak detection}
As with all Fourier filtering techniques, a crucial point is to properly define the masks for the continuous part (center peak), and oscillating parts (satellite peaks).
A basic approach is to define a fixed circular zone around the known position of the peak (band pass filtering).
This approach works very well for the continuous part since by definition it is always at the center of the Fourier plane, with a very limited radius (a few pixels).
However, as the satellite peaks grow in size with the increasing intensity and the non-linear dephasing, the area needs to be quite large, thus integrating noise.
Furthermore, if the angle of the reference beam changes, one needs to specify manually the position and radius of the circular mask.
In the following, we give the detailed procedure to automatize this crucial analysis step using the \texttt{scikit-image} Python package \cite{scikit-image}.

In order to identify the satellite peak, we use a common edge detection scheme which consists in working with the image gradient instead of the direct image.
For an initial intensity map $I(k_x, k_y)$, we work with the transformed  $I_{log}(k_x, k_y)=\text{log}(|\partial_{k_x} I(k_x,k_y)| + |\partial_{k_y} I(k_x,k_y)|)$.
This transformation allows to make sharp features such as the satellite peaks more apparent, while spreading the histogram of the highly contrasted Fourier transformed image.
Moreover, we fill the continuous part with the average value of the image in order to avoid detecting it with the feature detection procedure.

We then threshold and convert the image to an 8 bit integer type by eliminating all values below 60\% of the $I_{log}(k_x, k_y)$ maximum, to remove the noise floor of our image.
We then convert the image to binary using the Otsu thresholding function \texttt{threshold\_otsu} from skimage's \texttt{filters} module.
We remove small objects and holes, before detecting features using the \texttt{label} function from the \texttt{measure} module.
Finally, we select the largest area feature which is (empirically) always one of the two satellite peaks.
This last step allows to also measure the centroid of the satellite peak in order to use this information to demodulate (i.e removing the tilt induced by the reference beam). 

\noindent The general procedure described previously can be summarized as:
\begin{itemize}
    \item Log and gradient transform: \\$I_{log}=\text{log}(|\partial_{k_x} I(k_x, k_y)| + |\partial_{k_y} I(k_x,k_y)|)$.
    \item Thresholding to remove noise floor:\\ $I_{log} = I_{log}\left[I_{log}>0.6\cdot\text{max}(I_{log})\right]$.
    \item Binary thresholding: $\text{mask} = I_{log} > \text{Otsu}(I_{log})$.
    \item Erosion of small features and holes.
    \item Satellite peak detection and centroid measurement.
\end{itemize}
\paragraph{Practical considerations}
Two important conditions are required for the reconstruction to be successful: the reference beam needs to be collimated (in order to avoid diverging beam phase contributions), and the signal beam must be small enough that it does not fill the whole sensor such as to leave a border of low intensity to serve as a zero reference of the non-linear phase.
In practice these conditions are realized by taking a part of the signal beam as reference just before the non-linear medium with a polarized beam splitter.
This beam is then enlarged with a 4f telescope and recombined with the signal beam after the medium with a non polarized beam splitter (after having realigned the polarization). 
These conditions make this technique perfectly suited for measuring large non-linear dephasings.
Moreover, it is recommended when spatial resolution is needed (i.e a 2D map of the refractive index).
On the contrary, weak dephasings (below a few radians) are challenging to measure with this technique as they are comparable to the fluctuations due to convective currents, and therefore we present in the following a complementary phase retrieval method for weak non-linearity.

\begin{figure}[h]
    \centering
    \includegraphics[width=0.95\columnwidth]{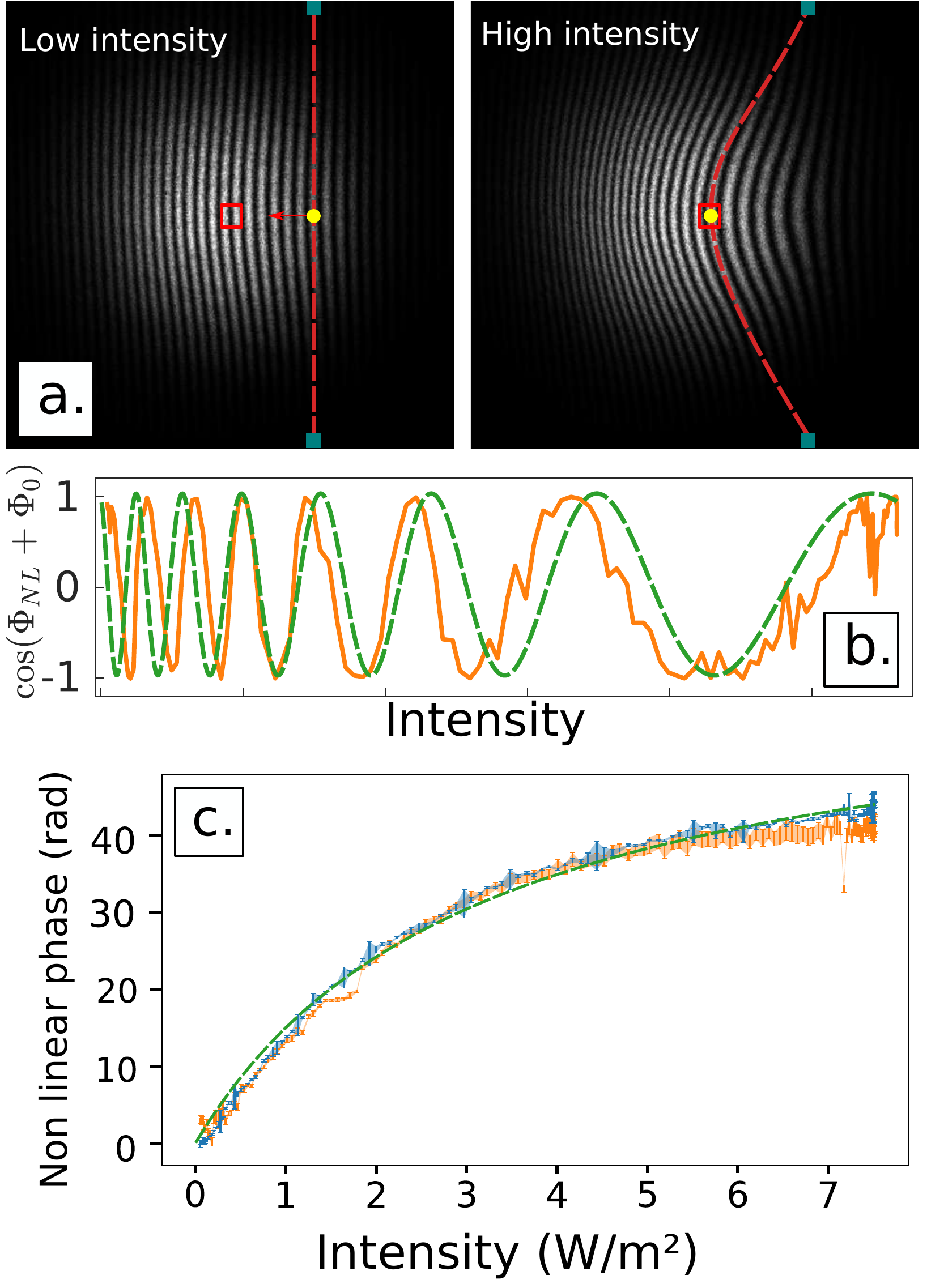}
    \caption{Non-linear phase measurement through direct monitoring of the interferogram for $\Delta$ = -4~GHz, beam waist $w_0$ = 1.85~mm, T=150°C and a maximal laser power P=560~mW.
    The cell used is an isotopically pure Rb$^{87}$ cell of 10~mm.
    a) The interference fringes bend and shift with increasing intensity as highlighted by the red dashed line.
    The non-linear index can be reconstructed directly from this pattern using one of the two methods presented in the text.
    b) For the bucket detector method, the intensity value inside of the red pixel square in a) is monitored and the oscillations are reconstructed on the orange line. The green dashed line is the theoretical fit from eq.\ref{eq:Dn}.
    c) Non-linear phase as function of the intensity.
    The blue curve is the phase recovered via Fourier filtering, the orange curve is the phase recovered from the bucket detector method and the green dashed curve is the $\Phi_{NL}=n_2 \frac{I}{1+\frac{I}{I_s}}$ fit.}
    \label{fig:murad2}
\end{figure}

\subsection{Bucket detector phase retrieval method}
It is possible to replace the camera sensor of the Fourier filtering procedure by a bucket detector (e.g. a photodiode) and still retrieve the non-linear phase.
This method is therefore an interesting alternative in the situation where the use of a camera is not possible. Moreover, as mentioned previously, this technique has a better sensitivity for measuring weak non-linear dephasings. 

\paragraph{Cosine fitting analysis} 
In this approach, we replace the spatially dependent intensity of the gaussian beam by a temporal intensity ramp and we "follow" the evolution of the non-linear dephasing during the ramp by monitoring the intensity of a small region of interest at the center of the beam.
The bucket detector can be for example: a photodiode sensor after an iris or a small subgroup of pixels (if we decide to still use a camera) as demonstrated on the panel a) of Fig.\ref{fig:murad2}. 
Provided the intensity ramp is slow enough with respect to the response time of the bucket detector, the signal at the center of the image will alternate between bright and dark values as the fringes shift due to dephasing. 
One can then fit a sinusoidal curve to recover the non-linear phase: $I\propto \text{cos}(\Phi_{NL}+\Phi_0)$.

The precision of this method is limited by the stability of the interferometer, which should be stable enough so that the fringes do not shift of more than one period over the measurement time.
As explained earlier, the presence of convective currents around the cell, due to the its heating elements, slightly blur the interferogram and it is preferable to have a long exposure time in order to average out these fluctuations (at the expense of a reduced contrast).
If these fluctuations exceed 2$\pi$ rads, the counting may integrate an extra fringe and falsify the final result.
Nevertheless, this method is more accurate for weak dephasings, and is not affected by the phase of the reference.

\subsection{Comparison of the two phase retrieval methods}
We have compared the two methods and verified that both methods recover the same non-linear dephasing as shown in Fig.~\ref{fig:murad2}: the divergence between the two curves is less than 2\% RMS.
The choice of the analysis methods mainly depends on the experimental constraints and requirements.
The bucket detector method is suited to measure weak dephasings (but requires about hundred of datapoints to be accurate) while the Fourier filtering method allows for recovering a spatially resolved $\Delta n$ from a single picture but is only suited to detect larger dephasings.
This is because the fluctuations due to convective currents around the cell are on the order of a few radians. 
When the non linear dephasing is on the order of these fluctuations, this considerately decreases the accuracy of the Fourier filtering method. 
This can be seen in fig.\ref{fig:murad2}c with the first three orange points over-estimated by 3 rads over the blue points.
On the contrary, due to the definition of a small region of interest employed in the bucket detector technique, these fluctuations are seen as a constant dephasing and thus shift the fringes around the center of the bin: these fluctuations are mostly erased during averaging as long as they do not exceed $\pi$ rads.

As a general guideline, we suggest to use the bucket detector method (with a camera subregion) for precise calibration of a system (especially for weak non-linearity, bellow a few radiants dephasing) and the Fourier filtering method for a quicker measurement of larger non-linear phase shift.


\subsection{Extension of non-linear Mach–Zehnder interferometer to measure the temporal response of a non-linear medium}
\label{section:time-resolved}
\begin{figure}[h]
    \centering
    \includegraphics[width=0.8\columnwidth]{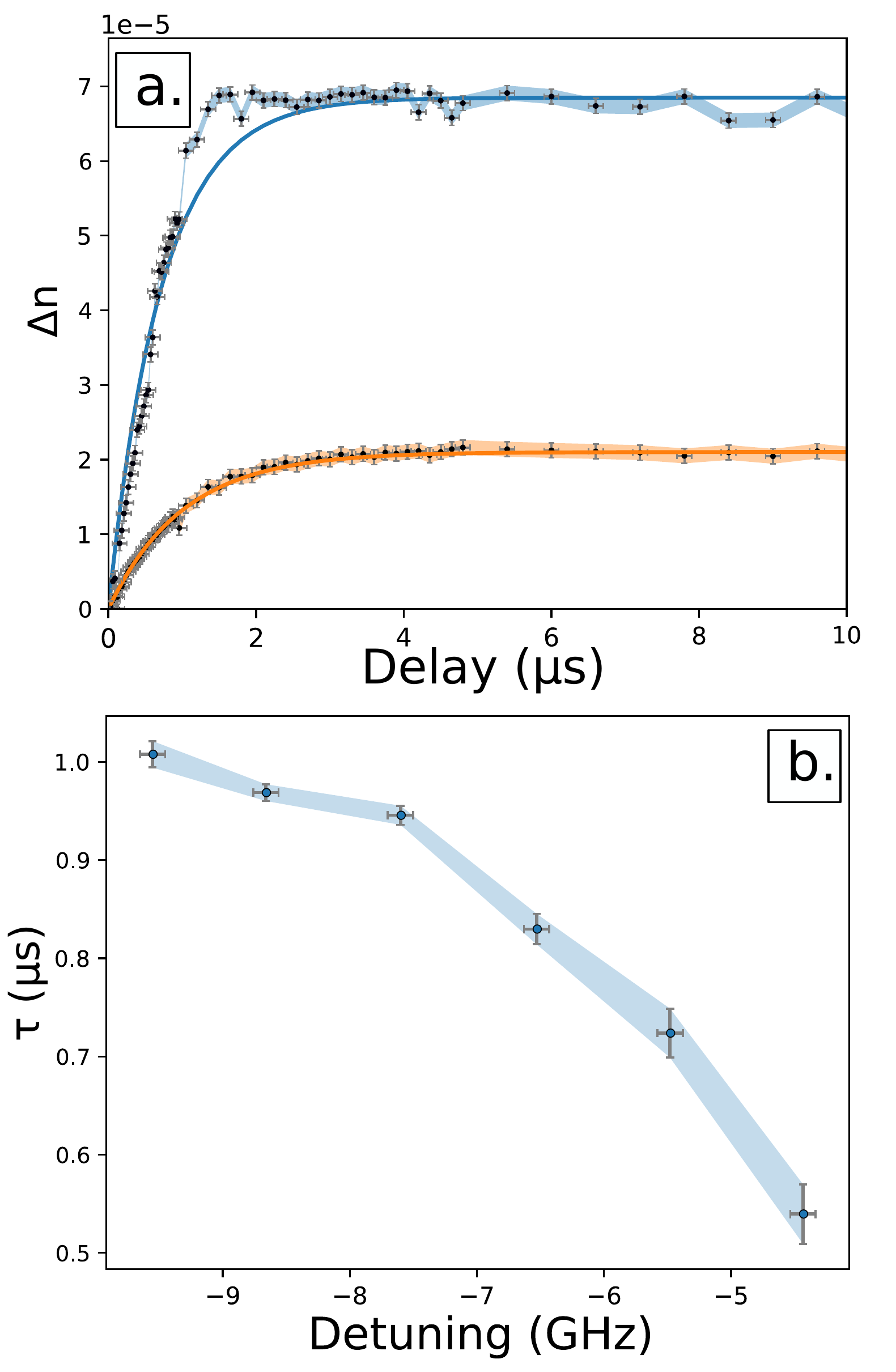}
    \caption{a) Non-linear index variation $\Delta n$ as function of the time delay between the signal and reference beams (crosses) for $w_0$=660~µm, T=140°C, P=400~mW and through a cell of isotopically pure Rb$^{87}$ of 10~cm.
    The continuous lines are a fit by $1-e^{-t/\tau}$. 
    The errorbars are highlighted with the shaded areas.
    The blue set is at $\Delta$ = -5.5 GHz, the orange set is at $\Delta$ = -9.5 GHz. 
    b) Fitted value of $\tau$ as function of the detuning $\Delta$.}
    \label{fig:pulse}
\end{figure}


Interestingly, our setup allows for time-resolved measurement of the non-linear index, not reported so far, for atomic vapors.
To demonstrate its potential for characterization, we apply this technique to retrieve the transient regime at short time for the non-linear index in a hot vapor of rubidium.

We study the temporal response of the non-linear medium by  using the interferometric method in a pulsed configuration.
The experimental setup remains unchanged, except that we now gate the reference and signal beams using an AOM as can be seen inside the dashed boxes of Fig.\ref{fig:setup}. 
We turn on the signal beam at a given time and tune the delay to gate the reference beam.
We therefore sample the non-linear index at a given delay after the signal has been switched on. 
The signal beam is between 50 and 20 $\mu$s long, while the reference was adjusted between 200 and 50~ns (which ultimately defined our temporal resolution) in order to keep a good contrast. Finally, our AOM could not produce pulses shorter than 30 ns.
We image the non-linear dephasing (essentially the non-linear response of the medium) at this delay.
For this experiment, we use the Fourier filtering phase retrieval method.

Experimental results are presented in Fig.\ref{fig:pulse} a) for two datasets at different detunings $\Delta$ from the signal laser with respect to the $D_2$ line of rubidium 87 (see Fig.~\ref{fig:waist} for the definition of $\Delta$).
The response of the medium is fitted by an exponential growth as $\Delta n(t) \propto 1-e^{-t/\tau}$ with a characteristic timescale $\tau$.
For the particular case of our atomic vapor medium, the value of $\tau$ can be related to the atomic structure of Rubidium.  
In the most basic approximation (commonly used for far detuned beam \cite{vsantic2018nonequilibrium}), rubidium can be considered as two-level atoms, and one could expect the time scale $\tau$ to be the lifetime of the excited state.
However, this approximation failed to describe quantitatively our results.
We refined this analysis by using a 3-level atomic model, as described in the supplementary material, using the eigenvalues of the evolution matrix of the density matrix.
This 3-level model gives a good quantitative estimate of $\tau$ that is consistent with the experiment within a fixed positive offset of 40\%.
In Fig.\ref{fig:pulse} b), we summarize our measurements of $\tau$ as function of $\Delta$: the closer we get to the resonance, the shorter the response time.
This effect can be quite large as switching from -6 to -5GHz nearly halves the response time from 0.8 to 0.55 µs. 
For this range of detunings, the relevant timescale before the medium reaches its steady state is below 1 $\mu$s.
This is of peculiar interest because this is comparable to the typical transit time of atoms in a beam of 1~mm at a temperature of 400K.
This opens interesting perspective for temporally non local interactions \cite{Vocke:15}.
These results highlight that the atomic evolution should not be described with the canonical steady-state approach and, in the next section,  we will consider the impact of transit effects on non-linear index measurements.




\section{Transit effects in hot atomic vapors}

In this last section, we use our method to investigate the effect of transit within finite transverse beams, on the non-linear index in hot atomic vapors.
One unavoidable aspect in all experimental setups is the finite size of the laser beams.
Surprisingly, the impact of this parameter on the non-linear index is not usually discussed.
For atomic vapors, this effect has dramatic consequences since it means that due to the thermal distribution of velocities and the random positions of the atoms, not all atoms will spend the same time in the beam. 
Slower atoms will spend a longer time in the beam and more quickly saturate to a steady state while faster atoms will interact only for a shorter time.
However, due to the Doppler shifting, faster atoms might be more resonant with the light and thus can be excited. 
Furthermore, as atoms come in and out of the beam, collision processes reshuffle the atoms internal state, and these collisions processes dictate at what speed the atoms will return (or not) in the beam as they govern the mean free path of the atoms in the vapor.
A complex combination of these effects is required to explain why (and how) the non-linear coefficient $n_2$ and the saturation $I_s$ vary with the beam diameter and it is difficult to form an intuition from these multiple parameters.
In the following, we present a general numerical method to study this transit effect and we compare it to our experimental results

\begin{figure}[h]
    \centering
    \includegraphics[width=0.95\columnwidth]{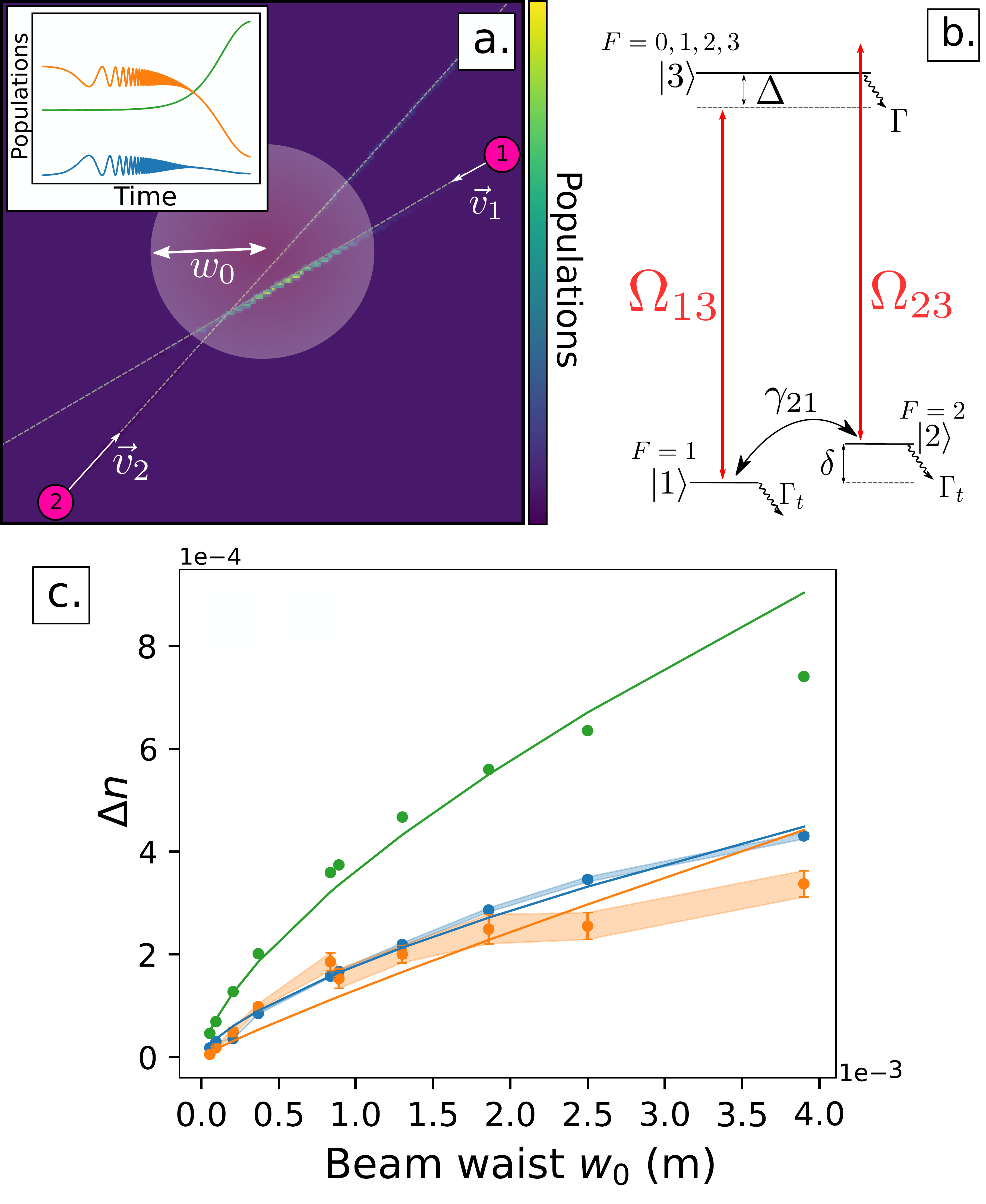}
    \caption{a.: Schematic view of the Monte-Carlo simulation: a beam of waist $w_0$ is represented in light pink. The atoms are initialized in their ground state far from the beam with random initial positions and velocities. As they traverse the computational window, the Maxwell-Bloch equations are numerically solved. 
    b.: The $D_2$ line of Rb$^{87}$ considered for the atoms. The laser field is shown in red coupling the two hyperfine ground states to the excited states manifold with two Rabi frequencies $\Omega_{13}$ and $\Omega_{23}$. 
    c.: the evolution of the non-linear index $\Delta n$ as a function of beam waist with detuning $\Delta$=~-2.2~GHz, a constant intensity I=17.8~W/cm$^2$, a cell of isotopically pure Rb$^{87}$ of 10~mm at a temperature of 150°C. The blue dots represent the numerical calculation, the orange dots the experimental data and the green dots the analytical calculation. The 1$\sigma$ confidence interval is indicated in the shaded areas. A linear fit on the curve allows to retrieve the power law exponent. }
    \label{fig:waist}
\end{figure}

\paragraph{Maxwell-Bloch equations}
The typical treatment of transit effects is to introduce a phenomenological transit rate into the optical Bloch equations \cite{glorieux2018quantum}. 
The atoms leaving the beam are seen as a loss of population in each of the ground states. 
The atomic structure of rubidium is depicted in Fig.\ref{fig:waist}. 
The excited states $\ket{3}$ (total momentum $F_e$=0, 1, 2, 3) in the hyperfine $5^{2}P_{\frac{3}{2}}$ manifold can decay either to two hyperfine ground states $\ket{2}$ ($F_g$=2) or $\ket{1}$ ($F_g$=1) in the $5^{2}S_{\frac{1}{2}}$ manifold, with total rate $\Gamma$=6.07 MHz \cite{Steck_numbers}. 
The dephasing rate between the states $\ket{1}$ and $\ket{2}$ is $\gamma_{21}$.
We consider the coupling of the laser field on each transitions and denote $\Omega_{ij}$ the corresponding Rabi frequencies.
The hyperfine splitting between $\ket{2}$ and $\ket{1}$ is noted $\delta$ and the detuning of the signal beam on the $\ket{1} \to \ket{3}$ transition is $\Delta$. The ground states splitting $\delta$ is 6.835 GHz, whereas the excited state manifold $5^{2}P_{\frac{3}{2}}$ total linewidth is 496 MHz, comparable to the Doppler linewidth of 400 MHz at 150°C. We thus ignore the hyperfine structure of the excited states. 
The resulting optical Bloch equations for the density matrix operator $\hat{\rho}$ describing the state can be found in the supplementary material eq.\ref{eq:mbe}.


\paragraph{Numerical simulation}

Instead of using this phenomenological approach, we propose to compute, ab-initio, the non-linear coefficient $n_2$ and the saturation $I_s$ without any phenomenological assumptions.
To do so  we proceed with a Monte-Carlo calculation.
In short, we set $\Gamma_t$ to zero, draw a set of random trajectories for individual atoms, solve the optical Bloch equation as function of time for each individual trajectory and sum over the atoms to obtain the average susceptibility over all trajectories.

The first step is to draw for each velocity class, a number $N_{traj}$ of random trajectories through the beam (typically $10^5$).
These trajectories are bounded to a discretized computation box of $N\times N$ pixels. 
We then solve the Maxwell-Bloch equations of the atomic state through the medium and average out the time dependence of the atomic response, at each point of the space grid, by accumulating the state of each passing atom. 
Each point of the grid is then averaged by the number of counts, and renormalized by the atomic density.

Solving the differential system is done  using a stiff ODE (Ordinary Differential Equations) solvers. 
Using a stiff solver allows to optimize the calculation time for problems, as ours, with different time scales.
Using a heavily optimized parallel framework provided by Julia's ODE library allows these calculations to run in a reasonable time.
However, care needs to be taken in the formulation of the right-hand side of the equation in order to take full advantage of this framework.
Fully optimized code to run these simulations and a discussion about design choices are provided in supplementary materials.

From these computations, we retrieve the coherence maps $\rho_{13}(\textbf{r})$ / $\rho_{23}(\textbf{r})$ and we calculate the electric susceptibility for this velocity class using the local electric field at each point $\mathcal{E}(\textbf{r})$ :
\begin{equation}\label{eq:chi}
\chi(\textbf{r})=\frac{2N}{\varepsilon_0\mathcal{E}(\textbf{r})}(\mu_{23}\rho_{23}(\textbf{r})+\mu_{13}\rho_{13}(\textbf{r})).
\end{equation}
We carry out the whole calculation for $N_v$ velocity classes (typically  20) and, finally, we compute the weighted average of the susceptibility against the velocity distribution at a given temperature. 

Comparing a high and a low intensity run allows us to recover the non-linear index $\Delta n$ in a thin slice of a rubidium vapor : the non linear index is the difference of the optical indices of high and low runs. 
We can refine the model in order to simulate the effect of absorption by doing these calculations for several intensities, representative of the intensities in the medium along the propagation (due to absorption, the intensity of the beam decreases exponentially). 
As the beam progresses inside of the cell, the medium becomes exponentially less saturated and the atomic response gets stronger. 
However as the intensity drops, the non-linear index change $\Delta n$ decreases.

\paragraph{Results and limitations}

The results of the simulations for the non-linear index $\Delta n$ as function of the beam waist diameter are shown in Fig.~\ref{fig:waist}~b).
With no adjustable parameters, our ab-initio model show a quantitative agreement down to 10\% .
Moreover, the behaviour against changes in waist size shows good qualitative agreement for the saturation behaviour of the medium: the fitted power law exponent for the experimental data is $0.84\pm0.09$ and for the computed results it is $0.77\pm 0.03$.
The analytical model presented in the supplementary material significantly oversetimates $\Delta n$ and shows a power law exponent of $0.67 \pm 0.03$.
Therefore, our ab initio numerical simulations without any assumptions on the transit rates and dephasing rates between the ground states improves on the phenomelogical approach.
\begin{figure}[h]
\includegraphics[width=0.7\linewidth]{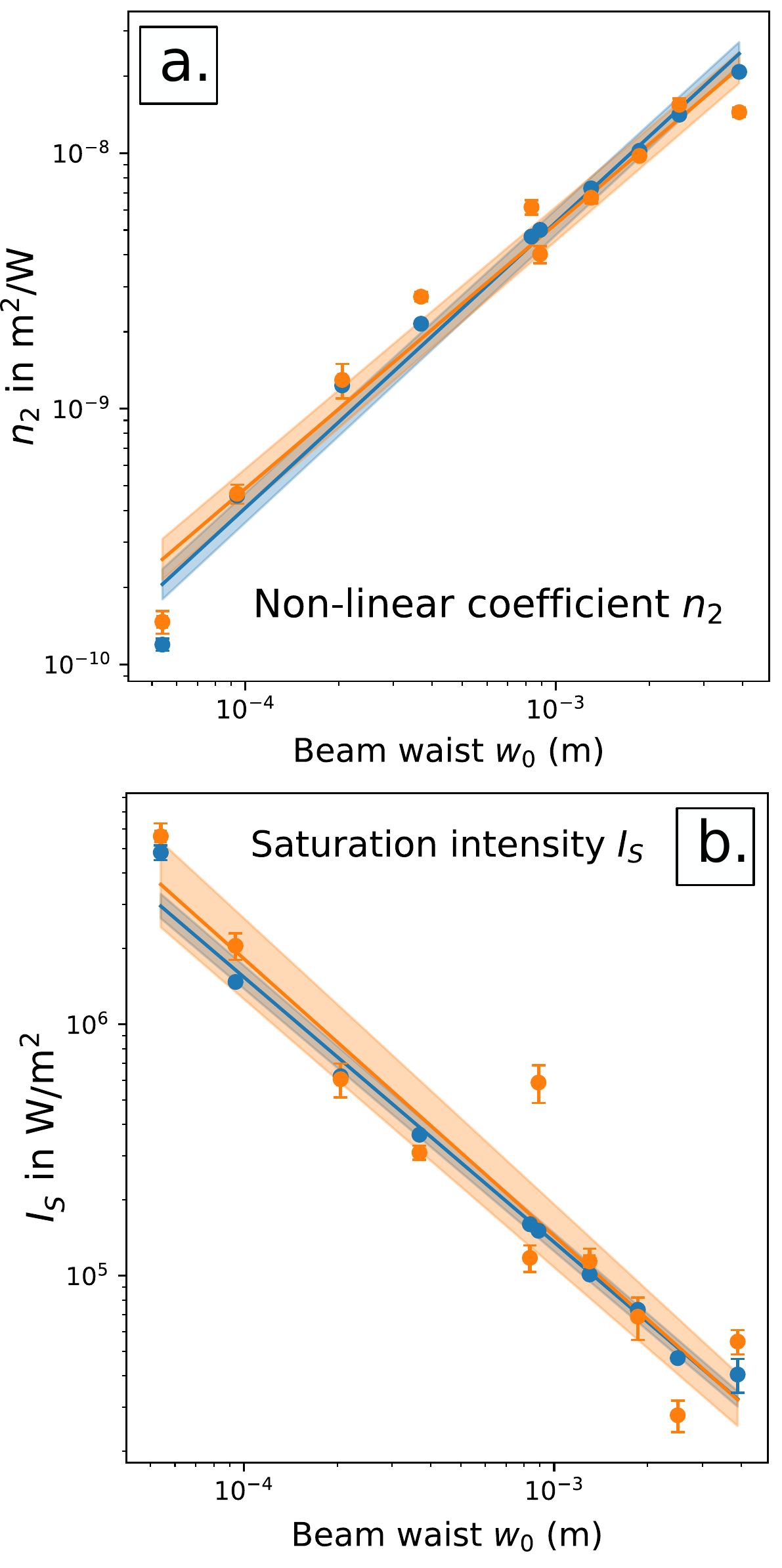}
\caption{a. : Variation of the non-linear coefficient $n_2$ with beam waist size. The blue dots are the numerical simulation, and the orange dots are the experimental points. The blue and orange lines represent the power law fit and the shaded areas the $1\sigma$ confidence interval. b. : Variation of the saturation intensity $I_S$ with beam waist size. The same color code as fig.\ref{fig:n2_Isat}.a applies.}
\label{fig:n2_Isat}
\end{figure}
By simulating the intensity ramps used for the measurement of $n_2$, one can also compare the behavior of $n_2$ and $I_S$ the saturation intensity versus waist size. The results of these simulations are shown in fig.~\ref{fig:n2_Isat}. We find excellent agreement between the fitted values of $n_2$ and $I_S$, and find that the non-linear coefficient increases linearly with waist size : the computed power law exponent $1.12 \pm 0.06$, while the experimental power law exponent is $1.03 \pm 0.08$. The saturation intensity decreases with the inverse of the beam size : the computed power law exponent is $-1.05 \pm 0.04$ while the experimental power law exponent is $-1.10 \pm 0.15$. 
  
The qualitative agreements of all curves seems to indicate that the transit effects are indeed the key ingredient for explaining the variation of $\Delta n$ with beam waist, and that three levels suffice to describe the non-linearities involved in our experiments.  

One could assume that the absence of inter-atomic collisions would impede on the results of the saturation intensity, especially at large beam waists. The dephasing collisions induce an additional rate between the two ground states that tend to reset the atomic medium thus increasing the saturation intensity.  However from the results presented above, they do not seem to play a role at the range of detunings we operate. 


\section{Conclusion}

We presented two complementary methods that allow for precise system characterization of optically thick samples with the possibility of a single shot, spatially and temporally resolved measurements.
We then investigated experimentally the spatial dependence of the non-linear coefficient $n_2$ with the change of beam size in a hot atomic vapor of rubidium.
Using Monte-Carlo simulations we showed that transit effects are fundamental to explaining the growth of $n_2$ with beam size.
This work can serve as a reference for measuring (or simulating) the non-linear index in hot atomic vapors.
It opens new perspective to better understand and measure all spatio-temporal effects in the behaviour of $n_2$, especially in fluid of light experiments, where transit effects and collisions lead to non-local interactions spatially and temporally \cite{skupin_nonlocal_2007,azam2021dissipation}.

\section*{Acknowledgments}
TA wrote the manuscript and performed analysis and numerical simulations, MA conceived and implemented the experimental methods, GB implemented the temporally resolved measurements, TB and TP participated to the experiments and data analysis, AB and QG wrote the manuscript and conceived the experiment.
The authors thank Q. Fontaine, T. Boulier, A. Urvoy and A. Sheremet for stimulating discussions.
We acknowledge financial support from the H2020-FETFLAG-2018-2020 project “PhoQuS” (n.820392).
QG and AB are members of the Institut Universitaire de France.

\bibliography{n2}
%
\section*{Appendices}
\appendix

\section{Maxwell-Bloch equations}
\begin{equation}
\partial_{t}\hat{\rho} = A\hat{\rho} + b 
\end{equation}
\begin{widetext}
\begin{equation}\label{eq:mbe}
\begin{split}
    A &=  \begin{pmatrix}
    -\Gamma_t -\frac{\Gamma}{2} & -\frac{\Gamma}{2} & 0 & 0 & \frac{i\Omega_{13}^{*}}{2} & -\frac{i\Omega_{13}}{2} & 0 & 0 \\
    -\frac{\Gamma}{2} & -\Gamma_t-\frac{\Gamma}{2} & 0 & 0 & 0 & 0 & \frac{i\Omega_{23}^{*}}{2} & -\frac{i\Omega_{23}}{2} \\
    0 & 0 & -\tilde{\gamma}_{21} & 0 & \frac{i\Omega_{23}^{*}}{2} & 0 & 0 & -\frac{i\Omega_{13}}{2} \\
    0 & 0 & 0 & -\tilde{\gamma}^{*}_{21} & 0 & -\frac{i\Omega_{23}}{2} & \frac{i\Omega_{13}^{*}}{2} & 0 \\
    i\Omega_{13} & \frac{i\Omega_{13}}{2} & \frac{i\Omega_{23}}{2} & 0 & -\tilde{\gamma}_{31} & 0 & 0 & 0 \\
    -i\Omega^{*}_{13} & -\frac{i\Omega^{*}_{13}}{2} & 0 & -\frac{i\Omega^{*}_{23}}{2} & 0 & -\tilde{\gamma}^{*}_{31} & 0 & 0 \\
    \frac{i\Omega_{23}}{2} & i\Omega_{23} & 0 & i\frac{\Omega_{13}}{2} & 0 & 0 & -\tilde{\gamma}_{32} & 0 \\
     -\frac{i\Omega^{*}_{23}}{2} & -i\Omega^{*}_{23} & -\frac{i\Omega^{*}_{13}}{2} & 0 & 0 & 0 & 0 & -\tilde{\gamma}_{32}^{*}
    \end{pmatrix}, \\
    b &= \begin{pmatrix}
    \frac{\Gamma}{2} + G_1\Gamma_t \\
    \frac{\Gamma}{2} + G_2\Gamma_t \\
    0 \\
    0 \\
    -\frac{i\Omega_{13}}{2} \\
    \frac{i\Omega^{*}_{13}}{2} \\
    -\frac{i\Omega_{23}}{2} \\
    \frac{i\Omega^{*}_{23}}{2}
    \end{pmatrix},\ 
    \hat{\rho} = \begin{pmatrix}
    \rho_{11} \\
    \rho_{22} \\
    \rho_{21} \\
    \rho_{12} \\
    \rho_{31} \\
    \rho_{13} \\
    \rho_{32} \\
    \rho_{23} 
    \end{pmatrix}
\end{split}
\end{equation}
\end{widetext}
with the following definitions: 
\begin{equation}
\begin{split}
    \tilde{\gamma}_{32} &= \Gamma - i\Delta, \\
    \tilde{\gamma}_{31} &= \Gamma - i(\Delta-\delta), \\
    \tilde{\gamma}_{21} &= \Gamma_t + i\delta, \\
    \Gamma_t &=  \frac{2}{\sqrt{\pi}}\frac{u}{w_0}.
\end{split}
\end{equation}
$\Omega_{ij} = \frac{\mu_{ij}\mathcal{E}}{\hbar}$ with dipole moment $\mu_{ij}$ and an applied field $\mathcal{E}$.  
The rates $\tilde{\gamma}_{ij}$ represent the effective dephasing rates between each state $\ket{i}$ and $\ket{j}$.
Finally, $\Gamma_t$ is a phenomenological transit rate calculated from $w_0$ the beam waist and $u=\sqrt{\frac{2k_B T}{m}}$ the most probable speed at a given temperature. 
$G_i$ are the ground state degeneracies : $G_1=\frac{3}{8}$ and $G_2=\frac{5}{8}$.
Note that the dephasing rate $\gamma_{21}$ is an effective rate due to transit as the $\ket{2}\rightarrow\ket{1}$ transition is forbidden. 
As atoms come and go through the beam, the coherence between the two ground states decays.
From this analytical model, one can derive steady state, far detuned ($\Delta\gg\Gamma$) approximations for the populations and coherences of the atoms, and one can then derive the susceptibility of the medium considering the $\ket{2}\rightarrow \ket{3}$ transition:
\begin{equation}\label{eq:chi}
\begin{split}
    \chi_{23} &= \sqrt{\frac{2b(1+a)}{1+b}}\frac{G_2 N(T)\mu_{23}^2}{\varepsilon_0\hbar\Gamma}\frac{i-\Delta/\Gamma}{1+(\frac{\Delta}{\Gamma})^2 + (\frac{\mathcal{E}}{\mathcal{E}_S})^2} \\
    a &= \frac{\frac{\Gamma}{2}}{\frac{\Gamma}{2} + \Gamma_t},~b = \frac{\Gamma_t}{\frac{\Gamma}{2} + \Gamma_t}, ~\mathcal{E}_S = \sqrt{\frac{2b(1+a)}{1+b}}\frac{\hbar\Gamma}{\mu_{23}}
\end{split}
\end{equation}

\paragraph{Time evolution constant}

Within the previously mentioned model, an upper bound for the atomic response time is the longest time scale of the atomic system i.e the smallest frequency directly obtained from the smallest eigenvalue of $A$ in Eq.~(\ref{eq:mbe}) \cite{glorieux2010double}. 
As the evolution of the system is described by a simple differential equation whose homogeneous solution at a certain time $t_0$ is:
\begin{equation}
    \hat{\rho}(t_0) = e^{\int_{0}^{t_0}dtA(t)}\hat{\rho}(0)~.
\end{equation}
If $P$ is the basis change matrix of $A$ to its diagonal form $D$, we can rewrite the previous equality as follows:
\begin{equation}
    \hat{\rho}(t_0) = P^{-1}e^{\int_{0}^{t_0}dtD(t)}P\hat{\rho}(0)~.
\end{equation}
We can then easily derive this standard inequality :

\begin{equation}
\begin{split}
    e^{\int_{0}^{t_0}\text{dt}\text{min}_{i}[\text{Re}(d_{i})](t)}||\hat{\rho}(0)|| \leq
    ||\hat{\rho}(t_0)|| \\
    \leq  e^{\int_{0}^{t_0}\text{dt}\text{max}_{i}[\text{Re}(d_{i})](t)}||\hat{\rho}(0)||~,
\end{split}
\end{equation}

where $d_i(t)$ ($i\in [0,7]$) are the eigenvalues of $A(t)$ at a time t.\\
\section{Numerical considerations about ODE solvers}
The stiff solvers used for the numerical calculations are mainly (\texttt{TRBDF2} or \texttt{KenCarp58}) from the Julia package \texttt{DifferentialEquations.jl} \cite{rackauckas2017differentialequations}. 
The \texttt{EnsembleProblem} class is the workhorse of the code.
It provides a convenient framework to run large numbers of realizations of the same differential equation with different paramaters and/or different starting values.
Great care was taken into formulating the right-hand side term (RHS) of eq.\ref{eq:mbe} in the fastest way possible. For this, we use several optimization techniques as doing in place calculations of the RHS, avoiding bound checks or using faster mathematical calculations through compiler options. \\
It is also critical to optimize the solver working point between  stability and efficiency i.e between the precision of the results, and the speed at which the computer provides said results.
We enforced stability by imposing a maximum time step of 0.1 µs which is consistent with the analysis of the temporal response presented in the main text: we want to appropriately sample the transient behavior of the atoms (approximately 1-2 µs).

\section{Code availability} The code is available through GitHub at the following address : \url{https://github.com/Quantum-Optics-LKB/Transit}. Please note that it is available \textbf{without any express or implied warranty}.


\end{document}